# An efficient broad-band mid-wave IR fiber optic light source: Design and performance simulation

A. Barh, S. Ghosh, R. K. Varshney, and B. P. Pal[*]

*Department of Physics, Indian Institute of Technology Delhi, New Delhi, 110016, India*
[*]*bppal@physics.iitd.ernet.in*

**Abstract:** Design of a mid-wave IR (MWIR) broad-band fiber-based light source exploiting four-wave mixing (FWM) in a meter long suitably designed highly nonlinear (NL) chalcogenide microstructured optical fiber (MOF) is reported. This superior FWM bandwidth (BW) was obtained through precise tailoring of the fiber's dispersion profile so as to realize positive quartic dispersion at the pump wavelength. We consider an Erbium ($Er^{3+}$) - doped continuous wave (CW) ZBLAN fiber laser emitting at 2.8 µm as the pump source with an average power of 5 W. Amplification factor as high as 25 dB is achievable in the 3 – 3.9 µm spectral range with average power conversion efficiency > 32%.

## 1. Introduction

In recent years, there has been a surge and continued interest to leverage on the huge development witnessed in fiber optic telecommunication to develop fibers and fiber-based devices suitable for mid-IR spectral region (2-10 µm). Emerging potential applications like non-destructive soft tissue ablation in medical diagnostics, monitoring of combustion flow and gas dynamics through molecular absorption spectroscopy, semiconductor processing (e.g. in-situ real time monitoring of plasma etch rates), and huge military applications in the mid-wave IR (MWIR) spanning 3 - 5 µm region have lately attracted a lot of research investments [1-2]. MWIR wavelength region is particularly important since a large number of molecules undergo strong characteristic vibration band transitions in this domain, which is also known as "molecular fingerprint regime" e.g. various hydrocarbons, hydrochlorides and commonly used solvents show strong absorption in the range of 3.2 – 3.6 µm [2]. Compact fiber-based light sources for MWIR would find wide scale military applications as it is a clean atmospheric window for high power transmission leading to applications in heat sinking missiles, IR counter-measures, and also thermal imaging for low power night vision in defense. Therefore it has become strategically important to develop an efficient light source in this wavelength region.

Chalcogenide glass (S-Se-Te)-based microstructured optical fibers (MOFs) have been considered potentially very suitable for the MWIR region due to certain special properties, which could be exploited to realize devices for applications in this wavelength range [3-5]. Studies on MOFs have shown that, waveguide dispersion in them dominates over material dispersion in determining the total dispersion of such fibers. This dispersion tailoring feature along with the relatively high Kerr nonlinearity in chalcogenide glasses (achievable $n_2$ being as high as 100 times larger than that of conventional silica fiber) [3], in them make these fibers extremely suitable for a number of applications like signal processing [6], all optical switching [7], supercontinuum generation [8-10], wavelength translation via FWM [11-13] etc. However, realization of low transmission losses in chalcogenide MOFs is still a challenge [6]. On the other hand, their chemical durability, glass transition temperature, strength, stability etc. can be improved by doping with As, Ge, Sb, Ga for drawing an optical fiber. At present their fabrication technology is well matured though expensive [6, 14-16].

In this paper, we have numerically designed an efficient, broad-band (covering 3 - 4 µm spectral domain) mid-IR light source by exploiting the degenerate four-wave mixing (D-FWM) process through the extraordinary linear and nonlinear (NL) properties of chalcogenide glass-based MOFs. In order to achieve broad and flat continuous spectrum around the targeted signal wavelength in the mid-IR, we have designed the fiber so as to obtain low anomalous dispersion ($\beta_2 \leq 0$) and positive fourth order GVD parameter ($\beta_4$) at pump wavelength ($\lambda_p$). Keeping these targets in mind, a broad-band chalcogenide fiber-based efficient light source for use in the MWIR wavelength regime (3 – 3.9 μm) has been numerically designed and reported here. This work should be of interest to those involved with chalcogenide fiber fabrication for its possible fabrication and utilization in order to realize MWIR devices.

## 2. Numerical Modeling

In optical fibers, several nonlinear phenomena could be exploited to generate new wavelength(s). Under certain conditions, however, FWM is the dominant nonlinear mechanism for generating new wavelengths, provided a certain phase-matching condition is satisfied [3, 17]. For our design purpose we will consider input pump power ($P_0$) levels to be below 5 W, which is considerably lower than the threshold for the onset of stimulated Raman and Brillouin scatterings in fibers shorter than 10 m [3]. Under D-FWM process, pump photons of frequency $\omega_p$ get converted into a signal photon ($\omega_s < \omega_p$) and an idler photon ($\omega_i > \omega_p$) according to the energy conservation relation ($2\omega_p = \omega_s + \omega_i$) where, subscripts s, i and p stands for signal, idler, and pump, respectively. For efficient mixing, it is important to recognize that the following phase matching condition is satisfied:

$$\Delta \kappa = \gamma P_0 + \Delta k_L \tag{1}$$

where $P_0$ is the input pump power, $\gamma$ is the well-known effective NL coefficient, and $\Delta k_L$ is the linear phase-mismatch term that is chromatic and inter-modal dispersion dependent and is given by [3]

$$\Delta k_L = \sum_{m=2,4,6...}^{\infty} 2\beta_m(\omega_P) \frac{\Omega_S^m}{m!} + \Delta k_W \tag{2}$$

where $\beta_m$ is the $m^{th}$ order GVD parameter, $\Omega_s$ is the frequency shift ($\Omega_s = \omega_p - \omega_s = \omega_i - \omega_p$), and $\Delta k_W$ is the phase mismatch term due to waveguide dispersion, which can be neglected in single-mode fibers [3]. Under CW pump condition in a highly NL fiber, the maximum $\Omega_s$ depends on both the magnitude and sign of GVD parameters. In one hand, positive $\beta_4$ leads to broad-band and flat gain where as negative $\beta_4$ reduces the flatness and BW of FWM. Thus higher order dispersion management is very crucial in such fiber designs. Considering up to fourth order dispersion, for positive $\beta_4$ and negative $\beta_2$, $\Omega_s$ can be approximated as

$$\Omega_s = \sqrt{\frac{6|\beta_2|}{\beta_4}\left(1 \pm \sqrt{1 - \frac{\beta_4 \gamma P_0}{3\beta_2^2}}\right)} \tag{3}$$

From Eq. (3), we can see that two sets of signal and idler are generated for this particular choice of GVD and NL parameters, i.e. for two signal wavelengths ($\lambda_s$), the phase matching becomes perfect. Therefore, we have to optimize the fiber and launching parameters such that the overall signal spectrum, generated around these two phase matching $\lambda_s$ become broad and flat with sufficient amplification. This positive $\beta_4$ value in the vicinity of low negative $\beta_2$ at the $\lambda_p$ could be achieved by suitably adjusting the core cladding index difference ($\Delta n$) in addition to optimizing the waveguide dispersion and hence multi-order dispersion management is feasible to engineer the FWM efficiency.

In our design calculation, first we have studied the D-FWM performance under lossless, undepleted pump condition, where pump power is only transferred to signal and idler wave. The launch of a weak idler along with the pump improves the FWM efficiency since stimulated FWM is employed in place of spontaneous FWM. It may be noted that here we are referring to the idler as an input field and the new wavelength generated in the mid-IR region as the signal. The peak amplification factor ($AF$) for the generated signal becomes

$$AF_S = P_{S,out}/P_{I,in} = (\gamma P_0/g)^2 \sinh^2(gL) \tag{4}$$

where $P_{S,out}$ is the peak signal power at the output, $P_{I,in}$ is the input idler power, $L$ is the interaction length and the D-FWM amplification coefficient $g$ is given by [3]

$$g = \sqrt{(\gamma P_0)^2 - (\Delta\kappa/2)^2} \tag{5}$$

In the next step, assuming quasi-CW conditions, we have studied the complex amplitudes $A_j(z)$ (j = p, i, s) and powers ($P_{out}$) variation along fiber length to study the effect of pump depletion and material loss by numerically solving the following three coupled amplitude equations [18]:

$$\frac{dA_p}{dz} = -\frac{\alpha_p A_p}{2} + \frac{in_2\omega_p}{c}\left[\left(f_{pp}|A_p|^2 + 2\sum_{k=i,s} f_{pk}|A_k|^2\right)A_p + 2f_{ppis}A_p^* A_i A_s e^{j\Delta k_L z}\right] \tag{6}$$

$$\frac{dA_i}{dz} = -\frac{\alpha_i A_i}{2} + \frac{in_2\omega_i}{c}\left[\left(f_{ii}|A_i|^2 + 2\sum_{k=p,s} f_{ik}|A_k|^2\right)A_i + f_{ispp}A_s^* A_p^2 e^{-j\Delta k_L z}\right] \tag{7}$$

$$\frac{dA_s}{dz} = -\frac{\alpha_s A_s}{2} + \frac{in_2\omega_s}{c}\left[\left(f_{ss}|A_s|^2 + 2\sum_{k=p,i} f_{sk}|A_k|^2\right)A_s + f_{sipp}A_i^* A_p^2 e^{-j\Delta k_L z}\right] \tag{8}$$

where $\alpha_j$ is the loss at the wavelength $\lambda_j$, $n_2$ is the NL index coefficient and $\Delta k_L$ is the linear phase mismatch defined in eq. (2). The overlap integrals ($f_{jk}$ and $f_{ijkl}$) are defined as

$$f_{jk} = \frac{\langle |F_j|^2 |F_k|^2\rangle}{\langle |F_j|^2\rangle\langle |F_k|^2\rangle} \qquad f_{ijkl} = \frac{\langle F_i^* F_j^* F_k F_l\rangle}{\left[\langle |F_i|^2\rangle\langle |F_j|^2\rangle\langle |F_k|^2\rangle\langle |F_l|^2\rangle\right]^{1/2}} \tag{9}$$

where $F_j(x,y)$ is the spatial distribution of the fiber mode in which the $j^{th}$ field propagates inside the fiber.

3. **Proposed Fiber Design**

To achieve such tailored application-specific fiber designs, we focus on an arsenic sulphide ($As_2S_3$)-based MOF geometry with a solid core and holey cladding, consisting of 4 rings of hexagonally arranged holes embedded in $As_2S_3$ matrix. In order to reduce $\Delta n$, we assume that borosilicate glass rod would fill the holes. Compatible thermal properties of the $As_2S_3$ and borosilicate glass should allow feasibility of fabrication of such a holey MOF [19]. The wavelength dependence of the linear refractive index of $As_2S_3$ and borosilicate glass has been incorporated through Sellmeier formula [19]. To suppress the modal loss, cutoff wavelength for 1$^{st}$ higher order mode need to be as low as possible (below $\lambda_p$). One way to achieve this is to maintain the hole-diameter to pitch ratio ($d/\Lambda$) of the MOF below 0.45. However, during the numerical optimization process we also realized that it is very difficult to simultaneously achieve a negative $\beta_2$ and a positive $\beta_4$ for this range of $d/\Lambda$; increment of $d/\Lambda$ is needed. Additionally, to enhance the magnitude of $\gamma$ and to minimize the confinement loss ($\alpha_c$), $A_{eff}$ should be as low as possible at both $\lambda_p$ and $\lambda_s$. To attain sufficient signal amplification over the MWIR regime of 3 ~ 3.9 μm, we have assumed commercially available CW $Er^{3+}$-doped ZBLAN fiber laser emitting at 2.8 μm [2] as the pump and confined our numerical study to pump power levels below 5 W to suppress other potential NL effects. Dispersion parameters as well as the modal field were calculated by using commercially available CUDOS® software along with MATLAB® for solving the coupled differential equations numerically. During optimization of the MOF structure, a strong interplay was observed between $\Omega_s$, parametric amplification, and $α_c$ with variations in $d$ and $Λ$. After optimization, a high amplification with a considerably low $α_c$ (0.01 to 0.001 dB/m) at the $λ_p$ and the generated $λ_s$ (up to 4 µm) was achieved for $d/Λ = 0.5$, $Λ = 2.5$ µm (shown in Fig. 1). Due to this tight confinement of light inside $As_2S_3$ core, the mode fields experience almost uniform nonlinearity during propagation. Wavelength dependence of dispersion coefficient ($D$) and GVD parameters $β_2$ and $β_4$ are shown in Figs. 2(a) and (b), respectively. Fig. 2(a) clearly indicates that $λ_{ZD}$ falls at 2.792 µm, which is slightly below the $λ_p$. At a fixed $P_0$, maximum amplification ($AF_{s,max}$) depends on $γ$ and length of the fiber ($L$) as it increases exponentially with "$γP_0L$". But BW is inversely proportional to this $L$ as long as $L \gg L_{NL}$ ($= (γP_0)^{-1}$). Thus to maintain large BW, $L$ should be relatively short at the cost of peak amplification.

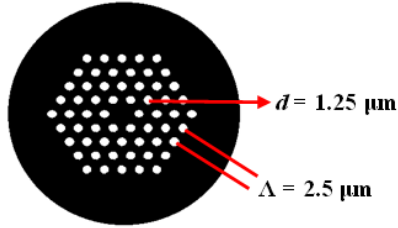

Fig. 1. Cross sectional view of the designed MOF. Cladding consists of 4 rings of borosilicate rods (white circles) embedded in the $As_2S_3$ matrix (black background). The diameter of the rod is $d$, and the centre to centre separation is denoted as pitch ($Λ$).

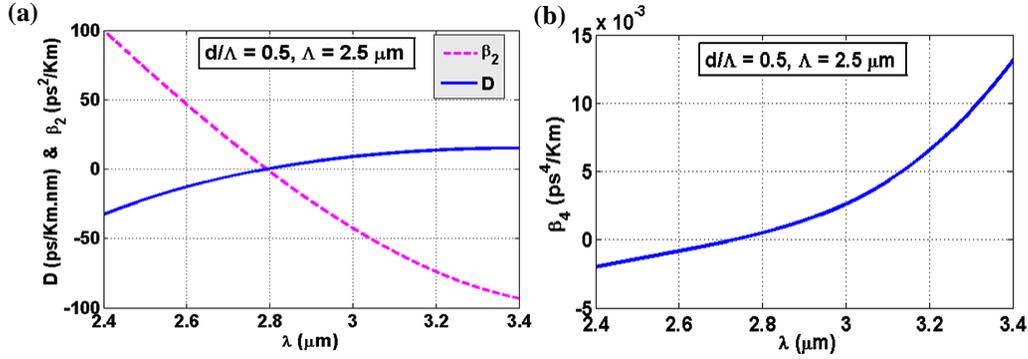

Fig. 2. Dispersion characteristics of $As_2S_3$ and Borosilicate based solid core MOF for $d/Λ = 0.5$ and $Λ = 2.5$ µm. (a) $D$ (blue solid curve) and $β_2$ (pink dashed curve) variation with operating wavelength ($λ$); $λ_{ZD} = 2.792$ µm. (b) Variation of $β_4$ with $λ$.

## 4. D-FWM performance under undepleted pump and lossless condition

During optimization of high-flat gain and maximum BW, a strong interplay was evident amongst $P_0$, $L$, and $λ_p$. If we detune $λ_p$ from $λ_{ZD}$, absolute value of $β_2$ increases leading to fluctuations in the output spectrum due to change in $Δκ$ around its zero value. This makes the spectrum less uniform as shown in Fig. 3(a); where only upper side of $λ_p$ is shown. In the vicinity of $λ_p$, the $AF_s$ decreases rapidly as $λ_s$ approaches $λ_p$. When the phase mismatch term becomes $γP_0$, the $AF_s$ becomes $(1+ γP_0 L)$ leading to a linear growth of signal from $λ_p$. From this figure, it can be interpreted that the best result could be achieved for $λ_p ≈ 2.797$ µm, where BW can be maximized. The GVD parameters $β_2$ and $β_4$ at this $λ_p$ were -1.20948 $ps^2$/Km and 3.71480 x $10^{-4}$ $ps^4$/Km, respectively. Calculated $A_{eff}$ at this wavelength came out to be quite small ~ 9.2 $µm^2$, which also helps in increasing the effective nonlinearity.

With 2.797 µm as the pump, and fixing $P_0$ at 5 W, we have studied the variation of gain spectrum for different values of fiber length (shown in Fig. 3(b)). This figure clearly indicates that the maximum $AF_s$ increases with increase in $L$ but at the cost of narrower BW. Thus to obtain high amplification of more than 35 dB optimum set of parameter were found to be $P_0$ = 5W, $L$ = 1 m, and $\lambda_p$ = 2.797 μm (shown in Fig. 3(b)). The achievable full width at half maxima (FWHM) is nearly 670 nm with confinement loss < 0.01 dB/m across the entire BW.

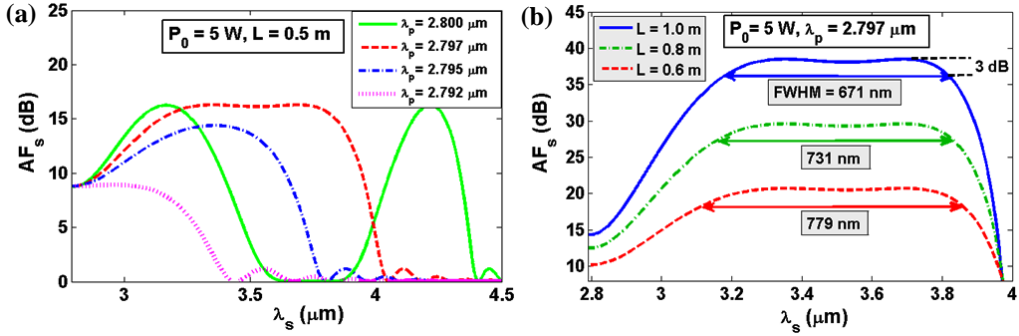

Fig. 3. (a) Variation of signal amplification factor ($AF_s$) for different $\lambda_p$ is shown. With $\lambda_p$ coinciding at $\lambda_{ZD}$ (= 2.792 μm), output signal spectrum is almost uniform around $\lambda_p$. With increase in $\lambda_p$ from $\lambda_{ZD}$, the BW as well as fluctuation increases. (b) Variation of $AF_s$ for a pump power of 5 W at 2.797 μm for different $L$ (0.6 – 1.0 m).

## 5. D-FWM performance under depleted pump including material loss

We now consider the material loss as reported earlier to be < 0.2 dB/m for the entire signal wavelength range [8]. We have fixed the designed fiber length ($L$) at 1 m, $P_0$ at 5 W and $\lambda_p$ at 2.797 μm. The generated two signal wavelengths for which the phase matching is almost perfect were 3.12 μm and 3.85 μm, and corresponding $\lambda_i$'s were 2.53 μm and 2.19 μm. Optimizing the spectral width as well as their phase matching $\lambda_s$ position, broad and flat spectrum can indeed be realizable.

To study the evolution of amplitudes ($A_j$), output powers ($P_{out}$), and $AF$ of pump, signal and idler along the propagation length ($L$), we numerically solved the three coupled amplitude equations (Eq. 6-8) for $P_0$ = 5 W, $\lambda_p$ = 2.797 μm, $\lambda_s$ = 3.85 μm and $P_{I,in}$ = 10 mW. Variations of $A_j$, $P_{out}$ and $AF$ are shown in Fig. 4(a), (b) and (c), respectively. From these figures we can see that even considering pump depletion and material loss, more than 1 W of output signal power is achievable.

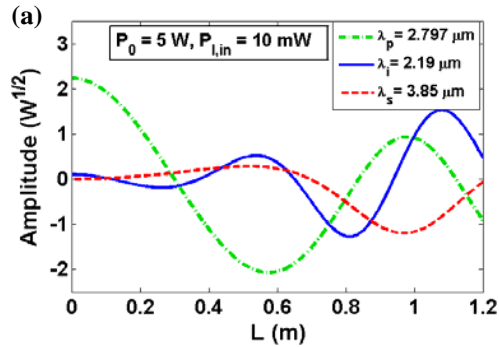

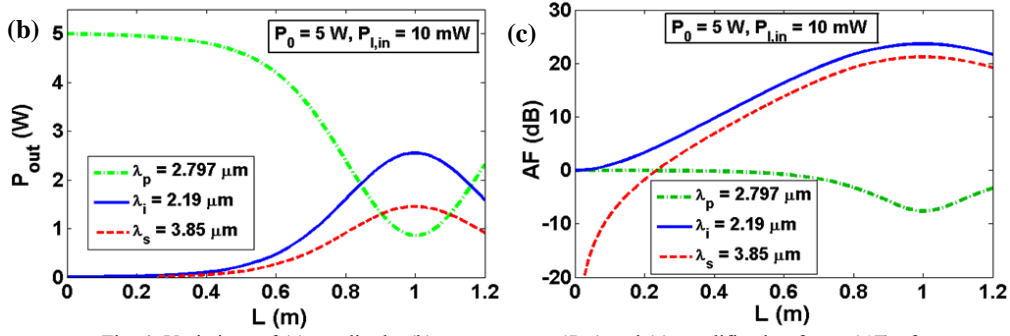

Fig. 4. Variations of (a) amplitude, (b) output power ($P_{out}$) and (c) amplification factor ($AF$) of pump, signal and idler with the fiber length ($L$) are shown. A weak idler of 10 mW at 2.19 μm is assumed along with 5 W of pump to initiate this D-FWM.

We have also optimized the $P_{I,in}$ to get maximum $P_{S,out}$ for 1 m of fiber length. These results are compiled in tabulated form in table 1, for $L = 1$ m, $P_0 = 5$ W, $\lambda_p = 2.797$ μm and $\lambda_s = 3.85$ μm, 3.12 μm. From this table it can be easily appreciated that to get a flat and broad output spectrum at the signal side, $P_{I,in}$ should lie between 8 to 11 mW. For 10 mW of $P_{I,in}$, the average output signal power over the entire output signal band is ≈ 1.64 W. Thus the power transfer efficiency ($P_{S,out}/P_0$) for this case is ~ 32.8 %, which is quite significant as a broad-band mid-IR light source. The entire output spectrum including both the idler and signal side is shown in Fig. 5(a) and (b), respectively, where for four different $P_{I,in}$, the variation of amplification factor is studied. Though the spectral BW is almost same like undepleted (i.e. loss less case), the maximum $AF_s$ decreases to ≈ 25 dB due to inclusion of pump depletion and spectral dependence of material loss. Such a fiber, if experimentally realized should be attractive as a mid-IR light source for a variety of applications outlined in the Introduction section.

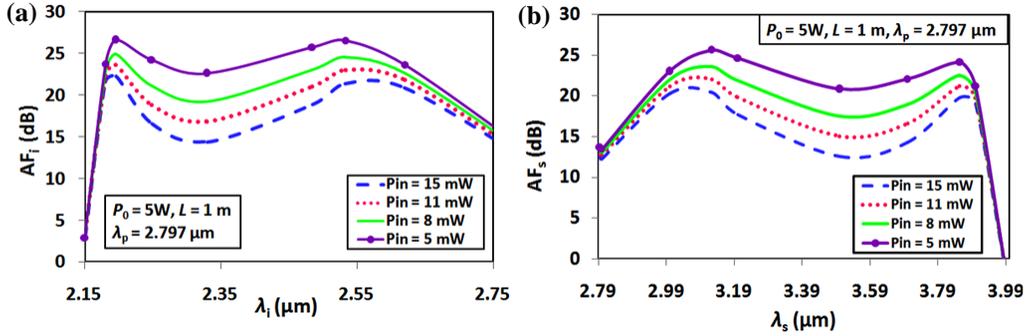

Fig. 5. Variation of amplification factor ($AF$) for different $P_{I,in}$ with pumping at 2.792 μm, $L = 1$ m and $P_0 = 5$ W. (a) The idler side of spectrum. (b) The signal side of spectrum. $AF_s$ as high as 25 dB is achievable, where overlap between two signal spectrum (one around 3.12 μm and another around 3.85 μm) makes the entire spectrum (3 – 3.9 μm) almost uniform.

Table 1. Variation of output signal power with input idler power

| $P_{I,in}$ (mW) | $P_{S,out}$ (W) ($\lambda_s = 3.85$ μm) | $P_{S,out}$ (W) ($\lambda_s = 3.12$ μm) |
|---|---|---|
| 20 | 1.3611 | 1.4135 |
| 15 | 1.4277 | 1.6595 |
| 12 | 1.4505 | 1.7356 |
| 11 | 1.4523 | 1.7724 |
| 10 | 1.4496 | 1.8189 |
| 8 | 1.4248 | 1.8376 |
| 5 | 1.2959 | 1.8314 |

## 6. Conclusions

We report a theoretical design of a broad-band MWIR light source by maximizing D-FWM band-width and efficiency in a highly nonlinear chalcogenide MOF. Through a detailed numerical and analytical study, for the first time to the best of our knowledge, we have shown that mid-IR power levels in excess of 1 W are achievable over the wavelength range of 3.1 – 3.9 μm with an amplification factor more than 25 dB for a 2.8 μm pump of 5W average power through a meter long specialty fiber based on our design. Additionally, a high power conversion efficiency (> 32%), and a very low confinement loss (< 0.01 dB/m) over the entire generated signal wavelength band should make our design route very attractive for making an all-fiber mid-IR light source. Potential application areas could be mid-IR spectroscopy, medical diagnostics, sensing, thermal imaging, astronomy and defense since the generated wavelength matches the second low-loss transparency window of the terrestrial atmosphere and the "fingerprint regime" for large number of molecules.

It would be interesting to undertake fabrication of chalcogenide fibers based on this design, though there could be several fabrication challenges like maintaining the required $d$, $\Lambda$ values throughout the fiber length, preparation of pure, low-loss, stable, stoichiometric glass compounds, etc. However, the important fact that there already exists well-matured fabrication technologies [6, 14-16, 20-22] to draw various chalcogenide MOFs, it should be of interest to invest efforts and money for fabrication of such application specific fibers where our methodology would serve an initial design platform. It is worthwhile to mention that, fabrication tolerance for pitch ($\Lambda$) and diameter of hole ($d$) with respect to their average value can be achieved within 2 to 4 % [22], which is below the tolerance limit of our designed fiber parameters for stable output as dispersion profiles remains nearly constant except the position of $\lambda_{ZD}$. In that case tunable pump (2.71 – 2.88 μm) [2] is needed to maintain the desirable GVD parameters at pump wavelength. Thus scope remains to improve this factor further.

### Acknowledgments

This work relates to Department of the Navy (USA) Grant N62909-10-1-7141 issued by Office of Naval Research Global. The United States Government has royalty-free license throughout the world in all copyrightable material contained herein. Some salient features of these results were recently reported by us at the international conference Photonics 2012 held at IIT Madras, Chennai, India. A.B. gratefully acknowledges support provided by CSIR (India) in the form of a Senior Research Fellowship award.

### References

1. A. Barh, S. Ghosh, G. P. Agrawal, R. K. Varshney, I. D. Aggarwal, B. P. Pal, "Design of an efficient mid-IR light source using chalcogenide holey fibers: A numerical study," J. of Optics (submitted).
2. S. D. Jackson, "Towards high-power mid-infrared emission from a fibre laser," Review Articles - Nature Photonics **6**, 423-431 (2012).
3. G. P. Agrawal, *Nonlinear Fiber Optics*, Academic, San Diego, Calif., (2007).
4. A. Zakery and S.R. Elliott, "Optical properties and applications of chalcogenide glasses: a review," J. Non-Cryst. Solids **330**, 1-12 (2003).
5. G. Boudebs, S. Cherukulappurath, M. Guignard, J. Troles, F. Smektala, F. Sanchez, "Linear optical characterization of chalcogenide glasses," Opt. Communications **230**, 331–336 (2004).
6. B. J. Eggleton, B. L. Davies, K. Richardson, "Chalcogenide photonics," Review Articles - Nature Photonics **5**, 141-148 (2011).
7. J. M. Harbold, F. Ö. Ilday, F. W. Wise, J. S. Sanghera, V. Q. Nguyen, L. B. Shaw, I. D. Aggarwal, "Highly nonlinear As–S–Se glasses for all-optical switching," Opt. Lett. **27**, 119-121 (2002).
8. J. Hu, C. R. Menyuk, L. B. Shaw, J. S. Sanghera, I. D. Aggarwal, "Computational study of 3-5 micron source created by using supercontinuum generation in $As_2S_3$ chalcogenide fibers with a pump at 2 micron," Opt. Lett. **35**, 2907-2909 (2010).

9. C. M. B. Cordeiro, W. J. Wadsworth, T. A. Birks, P. S. J. Russell, "Engineering the dispersion of tapered fibers for supercontinuum generation with a 1064 nm pump laser," Opt. Lett. **30**, 1980-1982 (2005).
10. D. I. Yeom, E. C. Mägi, M. R. E. Lamont, M. A. F. Roelens, L. B. Fu, B. J. Eggleton, "Low-threshold supercontinuum generation in highly nonlinear chalcogenide nanowires," Opt. Lett. **33**, 660-662 (2008).
11. M. R. E. Lamont, C. M. de Sterke, B. J. Eggleton, "Dispersion engineering of highly nonlinear $As_2S_3$ waveguides for parametric gain and wavelength conversion," Opt. Express **15**, 9458-9463 (2007).
12. C. S. Brès, S. Zlatanovic, A. O. J. Wiberg, S. Radic, "Continuous-wave four-wave mixing in cm-long Chalcogenide microstructured fiber," Opt. Express **19**, B621–B627 (2011).
13. J. D. Harvey, R. Leonhardt, S. Coen, G. K. L. Wong, J. C. Knight, W. J. Wadsworth, P. S. J. Russell, "Scalar modulation instability in the normal dispersion regime by use of a photonic crystal fiber," Opt. Lett. **28**, 2225-2227 (2003).
14. D. W. Hewak, "The promise of chalcogenides," Interview-Nature Photonics **5**, 474 (2011).
15. J. S. Sanghera, I. D. Aggarwal, L. B. Shaw, L. E. Busse, P. Thielen, V. Nguyen, P. Pureza, S. Bayya, F. Kung, "APPLICATIONS OF CHALCOGENIDE GLASS OPTICAL FIBERS AT NRL," J. of Optoelectronics and Ad. Materials **3**, 627-640 (2001).
16. M. El-Amraoui, G. Gadret, J. C. Jules, J. Fatome, C. Fortier, F. Désévédavy, I. Skripatchev, Y. Messaddeq, J. Troles, L. Brilland, W. Gao, T. Suzuki, Y. Ohishi, F. Smektala, "Microstructured chalcogenide optical fibers from $As_2S_3$ glass: towards new IR broadband sources," Opt. Express **18**, 26655-26665 (2010).
17. C. Lin, W. A. Reed, A. D. Pearson, H. T. Shang, "Phase matching in the minimum-chromatic dispersion region of single-mode fibers for stimulated four-photon mixing," Opt. Lett. **6**, 493-495 (1981).
18. G. Cappellini and S. Trillo, "Third-order three-wave mixing in single-mode fibers: exact solutions and spatial instability effects," J. Opt. Soc. Am. B **8**, 824- 838 (1991).
19. C. Chaudhari, T.Suzuki, Y. Ohishi, "Design of Zero Chromatic Dispersion Chalcogenide $As_2S_3$ Glass Nanofibers," J. of lightwave technol. **27**, 2095 – 2099 (2009).
20. J. S. Sanghera, C. Florea, L. Busse, B. Shaw, F. Miklos, I. D. Aggarwal, "Reduced Fresnel losses in chalcogenide fibers by using anti-reflective surface structures on fiber end faces," Opt. Express **18**, 26760-26768 (2010).
21. C. Quentin, B. Laurent, H. Patrick, N. T. Nam, C. Thierry, R. Gilles, M. Achille, F. Julien, S. Frédéric, P. Thierry, O. Hervé, S. Jean-Christophe, T. Johann, "Fabrication of low losses chalcogenide photonic crystal fibers by molding process," Proc. SPIE **7598**, 75980O-75980O-9 (2010).
22. F. Poletti, V. Finazzi, T. M. Monro, N. G. R. Broderick, V. Tse and D. J. Richardson, "Inverse design and fabrication tolerances of ultra-flattened dispersion holey fibers," Opt. Express **13**, 3728- 3736 (2005).